\documentclass[%
 aip,
 amsmath,amssymb,
 reprint,%
]{revtex4-2}

\usepackage{graphicx}% Include figure files
\usepackage{dcolumn}% Align table columns on decimal point
\usepackage{bm}% bold math
\usepackage[utf8]{inputenc}
\usepackage[T1]{fontenc}
\usepackage{mathptmx}

\begin{document}

\preprint{AIP/123-QED}

\title{Strong paramagnetic response in Y containing V$_{0.6}$Ti$_{0.4}$ superconductor}
% Force line breaks with \\

\author{SK Ramjan}
 %\affiliation[Also at ]{Physics Department, XYZ %University.}%Lines break automatically or can be forced with \\
 \affiliation{ 
 Free Electron Laser Utilization Laboratory, Raja Ramanna Centre for Advanced Technology, Indore 452 013, India%\\This line break forced with \textbackslash\textbackslash
}%
\affiliation{ 
Homi Bhabha National Institute, Training School Complex, Anushakti Nagar, Mumbai 400 094, India %\\This line break forced with \textbackslash\textbackslash
}%

\author{ L. S. Sharath Chandra}%
 \email{lsschandra@gmail.com}
 \affiliation{ 
 Free Electron Laser Utilization Laboratory, Raja Ramanna Centre for Advanced Technology, Indore 452 013, India%\\This line break forced with \textbackslash\textbackslash
}%
\author{Rashmi Singh}
%\affiliation[Also at ]{Physics Department, XYZ %University.}%Lines break automatically or can be forced with \\
\affiliation{ 
Laser and Functional Materials Division, Raja Ramanna Centre for Advanced Technology, Indore 452 013, India%\\This line break forced with \textbackslash\textbackslash
}%
\author{M. K. Chattopadhyay}
\affiliation{ 
 Free Electron Laser Utilization Laboratory, Raja Ramanna Centre for Advanced Technology, Indore 452 013, India%\\This line break forced with \textbackslash\textbackslash
}%
\affiliation{ 
Homi Bhabha National Institute, Training School Complex, Anushakti Nagar, Mumbai 400 094, India %\\This line break forced with \textbackslash\textbackslash
}

\date{\today}% It is always \today, today,

\begin{abstract}
We report here, the systematic field-cooled (FC) magnetisation of superconducting (V$_{0.6}$Ti$_{0.4}$)-Y alloys in presence of applied magnetic fields upto 7 T. Paramagnetic response is clearly observed just below the superconducting transition temperature (T$_{c}$) in low  fields ($\leq 0.2$~T). The lower Tc of the Y-rich precipitates as compared to the bulk, is the origin of flux compression and this leads to paramagnetic response. It is also observed that the magnetisation obtained during field cooled (FC) cooling cycle is lower than that of FC warming, for all the alloys in the field range 0.02-7 T. In addition, paramagnetic relaxation of FC moment is observed. We identify that these features of Y containing alloys are related to the high field paramagnetic Meissner effect (HFPME). Our analysis shows that the large difference in pinning strength of the different pinning centres generated due to Y addition to V$_{0.6}$Ti$_{0.4}$ alloy, is responsible for the observed effect. We provide further evidence to our claim in the form of extension of range in temperature and magnetic fields over which HFPME is observed when samples are subjected to cold work.
\end{abstract}

\maketitle

% oooooooooooooooooooooooooooooooooooooooooooooooooooooooooooooooooooo
\section{Introduction}

The superconducting V-Ti alloys have been the subject of interest in the last few decades due to its interesting physical properties such as negative temperature coefficient of resistivity~\cite{prekul1975effect}, presence of itinerant spin fluctuations \cite{prekul1975effect,matin2014influence,matin2014spin} and phonon dominating heat conduction in the superconducting state \cite{paul2019renormalization}. These alloys are highly machinable and are found to be promising candidates for high field applications, especially in the neutron irradiation environment \cite{tai2007superconducting, bellin1970critical}. 
\par Recently, a few V-Ti alloys in relatively high fields were found to show higher magnetisation in the field cooled warming (FCW) experiments than that in field-cooled cooling (FCC) experiments \cite{matin2013high,matin2015high}. Such behavior is also observed in many other superconductors such as Mo-Re \cite{sundar2015high},YBa$_{2}$Cu$_{3}$O$_{7-\delta}$  \cite{dias2016high,dias2001paramagnetic,dias2004paramagnetic}, Nb thin films\cite{terentiev1999paramagnetic} and is known as high field paramagnetic Meissner effect (HFPME). In some cases, the paramagnetic response is so large that, the field cooled (FC) magnetisation value is higher than the normal state magnetisation \cite{terentiev1999paramagnetic,prokhorov2009flux}. Relaxation of magnetisation towards more positive values is also a characteristic feature of HFPME \cite{matin2015high,dias2016high,terentiev1999paramagnetic}. The HFPME is observed in alloys in which anisotropic flux pinning exists and flux creep is significant\cite{matin2015high,matin2013high}. Paramagnetic response has also been observed in few systems in very low fields (mostly below lower critical field (H$_{c1}$) and is generally termed as Wohlleben effect \cite{svedlindh1989anti, kostic1996paramagnetic, ge2017paramagnetic,chu2006extrinsic, terentiev1999paramagnetic}.Various models of flux trapping and flux compression, formation of giant vortex state, nodes in the superconducting gaps and occurrence of $\pi$-junctions between superconducting grains are used to explain the Wohlleben effect \cite{chu2006extrinsic,svedlindh1989anti, kostic1996paramagnetic, ge2017paramagnetic,chu2006extrinsic, terentiev1999paramagnetic}.

\par Our previous studies show that inhomogeneous flux pinning along with significant flux creep are the reasons behind HFPME in the Ti-V alloys \cite{matin2013high, matin2015high}, and the absence of this phenomenon in V$_{0.6}$Ti$_{0.4}$ alloy is due to its low flux creep \cite{matin2015high}. Recently we have shown that, the rare-earths (RE) are immiscible in the  V$_{0.6}$Ti$_{0.4}$ phase and precipitate along the grain boundaries \cite{paul2021grain,ramjan}. It is also observed that the distribution of size of precipitates depends on the amount of RE element present in the alloy \cite{ramjan}. Moreover, the stress field of the precipitates can generate  dislocations in any sample. These mixed microsturcture is also responsible for the enhancement of critical current density in these alloys. The pinning strength of different kinds of defects depend on the temperature and field. Therefore, an anisotropic distribution of pinning may be expected. Such a situation is favourable to for the HFPME. This motivated us to study HFPME in V$_{0.6}$Ti$_{0.4}$-Y alloys.

\par In this paper we show that the all V$_{0.6}$Ti$_{0.4}$-Y  alloys exhibit paramagnetic Meissner effect (PME). The Y precipitates in the V$_{0.6}$Ti$_{0.4}$ alloys, and reduces the grain size. These Y-rich precipitates becomes weakly superconducting due to the proximity effect. We also show that the landscape of strength of pinning centres generated due to Y addition, is responsible for the PME at high as well as low fields. The flux compression inside the Y-rich precipitates, due to its lower Tc than the matrix is the origin of low field paramagnetic response.

%oooooooooooooooooooooooooooooooooooooooooooooooooooo
\section{Experimental Details}
The series of V$_{0.6}$Ti$_{0.4}$-Y samples were synthesized by melting high purity constituent elements in an arc melting furnace filled with pure Ar. The samples were flipped and remelted 5 times to ensure homogeneous mixing. The samples will be referred to by their short names as presented in table 1. The magnetisation measurements were performed on a portion of the samples, using Superconducting Quantum Interference Device based Vibrating Sample Magnetometer (MPMS-3 SQUID VSM, Quantum Design, USA). The magnetisation as a function of temperature at various fields is measured in the zero field cooling (ZFC), FCC and FCW protocols. In the ZFC protocol, the magnetic field is switched on after allowing the sample to cool down from above T$_{c}$ to 2 K, and the measurements are done while warming. After the sample reaches T $>$ Tc, the measurements are continued while cooling  down in the same applied field to 2 K (FCC). Subsequently, magnetisation is measured while warming up the sample (FCW).
\begin{table}[ht]
 \caption{Short names used for the samples as per their yttrium content.}
 \centering % used for centering table
 \begin{tabular}{c c} % centered columns (4 columns)
 \hline\hline %inserts double horizontal lines
 Elemental composition  & Short name \\ % inserts table
 %of pins & type &  form &  of maximum \\ [0.8ex] 
 %heading
 \hline % inserts single horizontal line
 V$_{0.6}$Ti$_{0.4}$ & Y0  \\ % inserting body of the table
 V$_{0.59}$Ti$_{0.4}$Y$_{0.01}$ & Y1 \\
 V$_{0.58}$Ti$_{0.4}$Y$_{0.02}$ & Y2 \\ 
 V$_{0.57}$Ti$_{0.4}$Y$_{0.03}$ & Y3 \\
 V$_{0.55}$Ti$_{0.4}$Y$_{0.05}$ & Y5 \\
 
 \hline\hline %inserts single line
 \end{tabular}
 \label{I} % is used to refer this table in the text
 \end{table}
The series of V$_{0.6}$Ti$_{0.4}$-Y samples were synthesized by melting high purity constituent elements in an arc melting furnace filled with pure Ar. The samples were flipped and remelted 5 times to ensure homogeneous mixing. The samples will be referred to by their short names as presented in table 1. The magnetisation measurements were performed on a portion of the samples, using Superconducting Quantum Interference Device based Vibrating Sample Magnetometer (MPMS-3 SQUID VSM, Quantum Design, USA). The magnetisation as a function of temperature at various fields is measured in the zero field cooling (ZFC), FCC and FCW protocols. In the ZFC protocol, the magnetic field is switched on after allowing the sample to cool down from above T$_{c}$ to 2 K, and the measurements are done while warming. After the sample reaches T $>$ Tc, the measurements are continued while cooling  down in the same applied field to 2 K (FCC). Subsequently, magnetisation is measured while warming up the sample (FCW).

\par High resolution images (2000X) of all the yttrium substituted samples were taken using Sigma Carl-Zeiss Scanning Electron Microscope (SEM). The presence of yttrium precipitates were confirmed and its distribution in the V-Ti matrix  was observed by Energy dispersive X-ray (EDX) analysis integrated with the SEM. A set of unetched and well-polished samples were used for EDX analysis, since etching of samples resulted in removal of yttrium and formation of pits.

\section{Results and discussion}
%....................Microstructural studies......................

\subsection{Microscopic features}

%...............Figure SEM/Meta...................

\begin{figure}[h]
\centering
  % El fichero es un eps y se   automaticamente a pdf con eps2pdf package
  \includegraphics[width=\columnwidth]{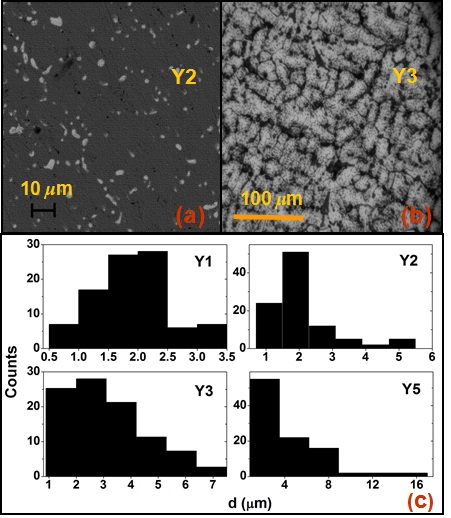}
\caption{(a) SEM image of Y2 alloy highlighting the Y-rich precipitates (white color) in Y2 alloy.(c) Optical mettalographic image of Y3 alloy revealing the grain boundaries and dislocations.(c) Histogram for the sizes of yttrium precipitates for all the alloys .}\label{1}
\end{figure}
%........................................

\par The detailed microscopic studies on Y0 and Y1-Y5 alloys can be found elsewhere \cite{matin2013magnetic, matin2013high}. Here, we highlight some important points required for this study. Fig.\ref{1}(a) shows the SEM image for Y2 alloy with 2000X magnification. The secondary phase particles dispersed in the V-Ti matrix , are the Y-rich precipitates(appearing in white color). Fig.\ref{1}(b) shows the optical image of chemically etched Y3 alloy.The grain boundaries are clearly revealed in form of dark boundaries. The grain sizes of Y1-Y5 were found to be in the range 1-50 $\mu$m \cite{ramjan}, which is an order less than the Ti-V alloys \cite{matin2015critical}. The grain size initially decreases with the increase in yttrium content up to 2 at. $\%$. On further addition of Y, the grain size starts to increase. Majority of Y-rich precipitates appearing as black dots, are found along the grain boundaries. The network of very small dots found inside the grains represents the dislocations. Fig.\ref{1}(c) shows the variation in size of the precipitates in the alloys Y1-Y5. The precipitates size varies from 0.4 $\mu$m in Y1 alloy to about 18 $\mu$m in Y5 alloy. Moreover, few precipitates of size 30 $\mu$m have also been found in Y5. The mean size of yttrium precipitates increases with yttrium concentration upto 3 at.$\%$. On further increase in yttrium concentration, the mean size of precipitates increases sharply due to liquid immscibility \cite{ramjan}. The number of precipitates is found to be maximum in Y2. For, higher yttrium concentration,  the increase in precipitate size results in decrease in their number.

% ................Flux compression.............

\subsection{Flux compression in Y samples}

%...........Fig.......low field PME............

\begin{figure}[h]
\centering
  % El fichero es un eps y se convierte automaticamente a pdf con eps2pdf package
  \includegraphics[width=\columnwidth]{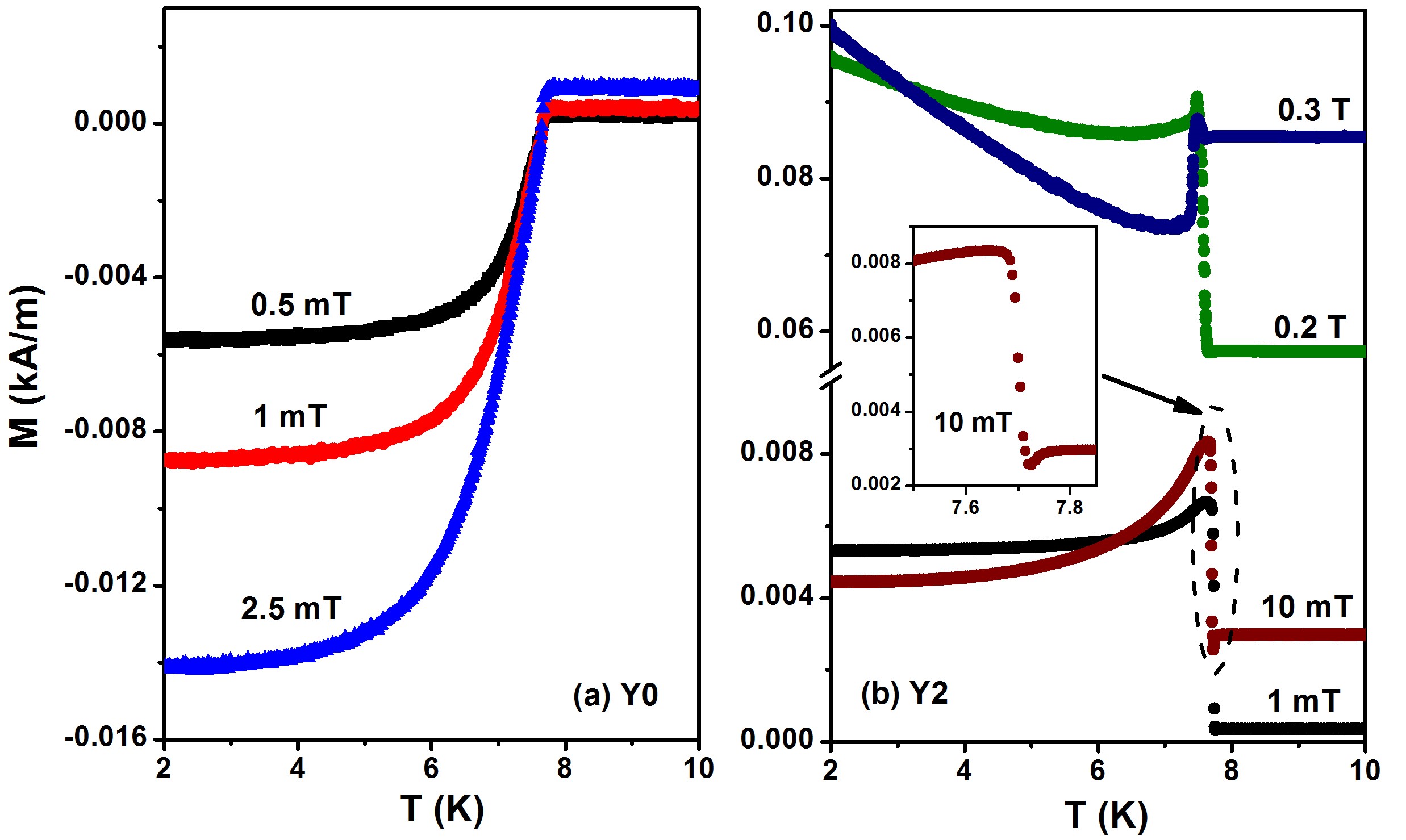}
  \caption{FCC (T) curves obtained in presence of various fields for a)Y0 b) Y2 alloys. The positive magnetisation signal is clearly in Y2 alloy. The inset to the fig.b shows the close-up view near the Tc in 0.01 T.  }
  \label{2} 
\end{figure}

The FC magnetisation curves in very low fields is shown for Y0 and Y2, in fig.\ref{2}(a) and\ref{2}(b) respectively. The positive magnetisation signal is clearly observed in case of Y2 alloy in the field range 0.001-0.2~T. While cooling down through the superconducting critical temperature (T$_{c}$), an initial dip  is observed, followed by a rise in the magnetisation above the normal state value. This is clearly evident from the inset provided in fig.\ref{2}(b), and is a clear indication of PME. The sudden rise in M just below Tc is observed in field values upto 0.2 T. Similar effect has been observed in multiphase In-Sn alloys, where two different phases superconduct at different temperatures \cite{chu2006extrinsic}. The In$_{10}$Sn$_{90}$ alloy consist of $\beta$-Sn phase (T$_{c2}$ = 3.7 K) encapsulated in the $\gamma$-InSn (T$_{c1}$ = 4.5 K) matrix. When the alloy is cooled  below T$_{c1}$, the  $\gamma$-InSn matrix first excludes the flux lines. This gives rise to the initial diamagnetic response observed. On further cooling, the flux lines gets trapped and subsequently compressed inside the $\beta$-Sn particles. In similar lines, we can expect flux compression in the V$_{0.6}$Ti$_{0.4}$-Y alloys provided, the Y-rich precipitates are superconducting.  Moreover, no signatures of PME is observed in case of Y0 alloy. This indicates the effect might be due to the presence of yttrium precipitates. 
%.............Y50V-Ti50 image..............
\begin{figure}[ht]
\centering
  % El fichero es un eps y se convierte automaticamente a pdf con eps2pdf package
  \includegraphics[width=\columnwidth]{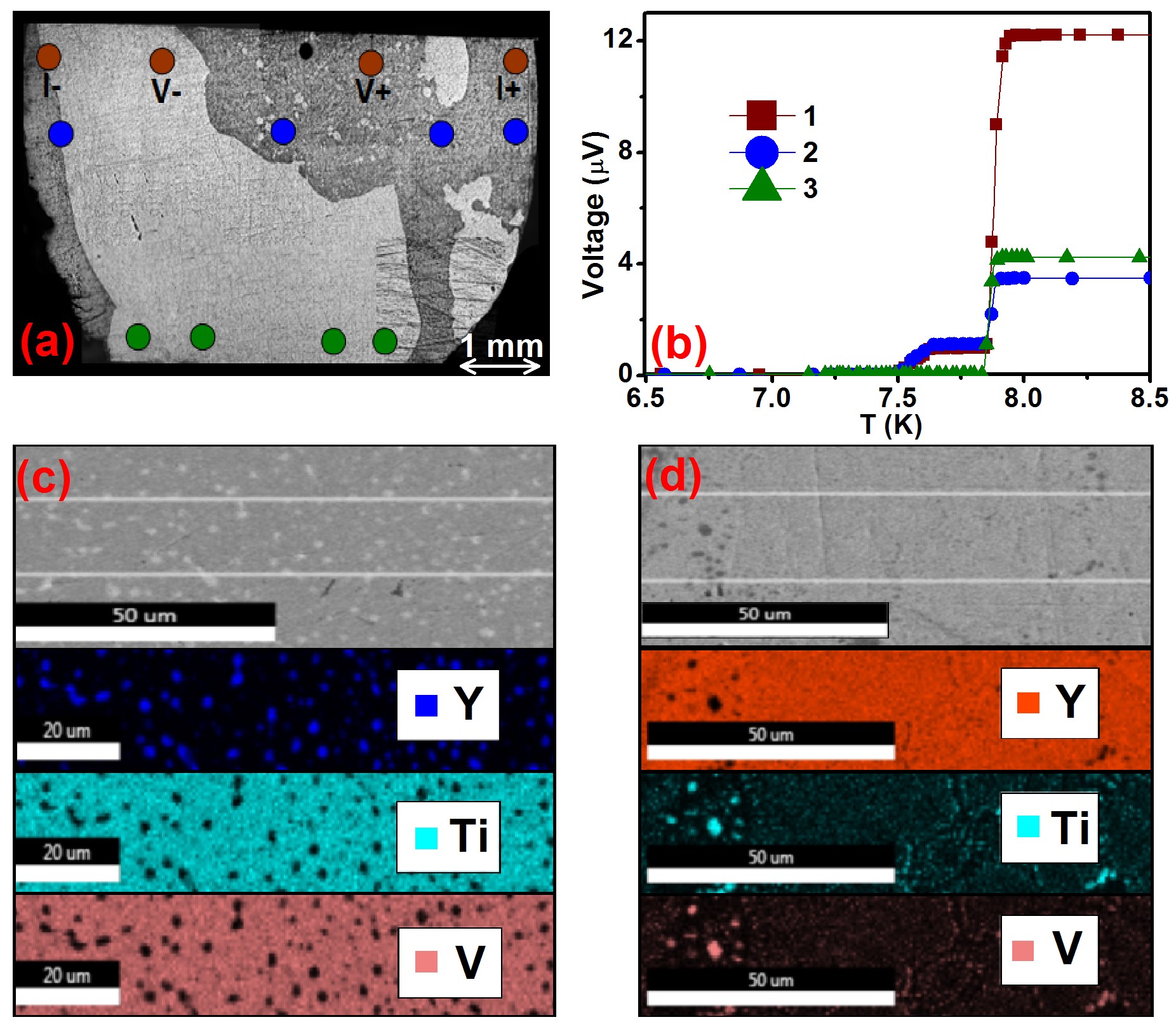}
  \caption{(a) Optical mettalography image of (V$_{0.6}$Ti$_{0.4}$)$_{0.5}$Y$_{0.5}$ alloy showing two different phases. The solid circles represent various contact points used for resistivity measurements. (b) Measured voltage as a function of temperature in 3 different configurations for the voltage and current leads from fig.(a). (c) and (d) shows the SEM image and elemental analysis of a region from bright and dark phase (of fig.(a)), respectively. The at. conc. of Y in region considered in (c) an (d) is 5$\%$ and 84$\%$ respectively. }
  \label{3} 
\end{figure}
\par Elemental Y under ambient pressure is non superconducting upto lowest temperature, due to the effect of spin fluctuation \cite{bose2008electron}. However, existence of superconductivity at higher pressures, infers that yttrium has a tendency to superconduct \cite{hamlin2006superconductivity}. In order to explore whether the Y-rich precipitates are superconducting, (V$_{0.6}$Ti$_{0.4}$)$_{0.5}$Y$_{0.5}$  sample with higher Y concentration was prepared. The optical micrograph of a slice of the sample is shown in fig. \ref{3}(a). The sample was found to have two phases, which can be understood from the phase diagram provided in reference \cite{chan2010thermodynamic}. The elemental analysis of a portion each from the bright and dark regions of fig.\ref{3}(a), is shown in fig.\ref{3}(c) and \ref{3}(d) respectively. The bright portion represents V$_{0.6}$Ti$_{0.4}$-rich phase with only 5 at.$\%$Y(similar to as obtained in Y1-Y5 alloys \cite{ramjan}). The darker regions are Y-rich (see fig.\ref{3}(d)), with Y concentration upto 84 at.$\%$. Very few pockets of V$_{0.6}$Ti$_{0.4}$ rich regions are observed in fig.\ref{3}(d), without forming a continuous path. The results of the resistivity measurements carried out with three different configuration of voltage leads, are shown in fig.\ref{3}(b). For configuration 2 and 3, both the voltage leads were placed in Y-rich and V$_{0.6}$Ti$_{0.4}$-rich phase, respectively. Measurements in configuration 1 were done by placing one voltage lead each in two different phases. A single step superconducting transition at 7.8 K, was observed for configuration 3. This temperature matches with the T$_{c}$ obtained from resistivity for Y5 alloy \cite{ramjan}. However, configuration 2 shows a two step transition before reaching zero resistivity. The initial drop is observed at the 7.8 K, which is due to the presence of V$_{0.6}$Ti$_{0.4}$-rich phase. However, the amount of Y0 rich phase present in Y-rich region is below the percolation threshold. Hence, zero resistivity is not observed at 7.8 K in configuration 2. On further decreasing the temperature, a gradual decrease to zero resistivity is obtained below  7.6 K. This indicates, that the Y-rich region becomes superconducting at a lower temperature due to the proximity effect induced by the V$_{0.6}$Ti$_{0.4}$-rich phase. In case of configuration 3, two step superconducting transition is observed as in case of configuration 1.
\par From above experimental evidences, it is possible that a major portion of the Y-rich precipitates are superconducting at a temperature below the bulk T$_{c}$ in Y1-Y5 samples. This leads to flux compression, and the observed paramagnetic response in V$_{0.6}$Ti$_{0.4}$-Y alloys. The observation of this effect upto 0.2 T can be understood considering the weak nature of superconductivity induced by proximity effect. The application of small field, renders the Y-rich precipitates normal. Hence, we do not observe the flux compression above 0.2 T.

\subsection{High field paramagnetic Meissner effect}

\begin{figure}[t]
\centering
  % El fichero es un eps y se convierte automaticamente a pdf con eps2pdf package
  \includegraphics[width=\columnwidth]{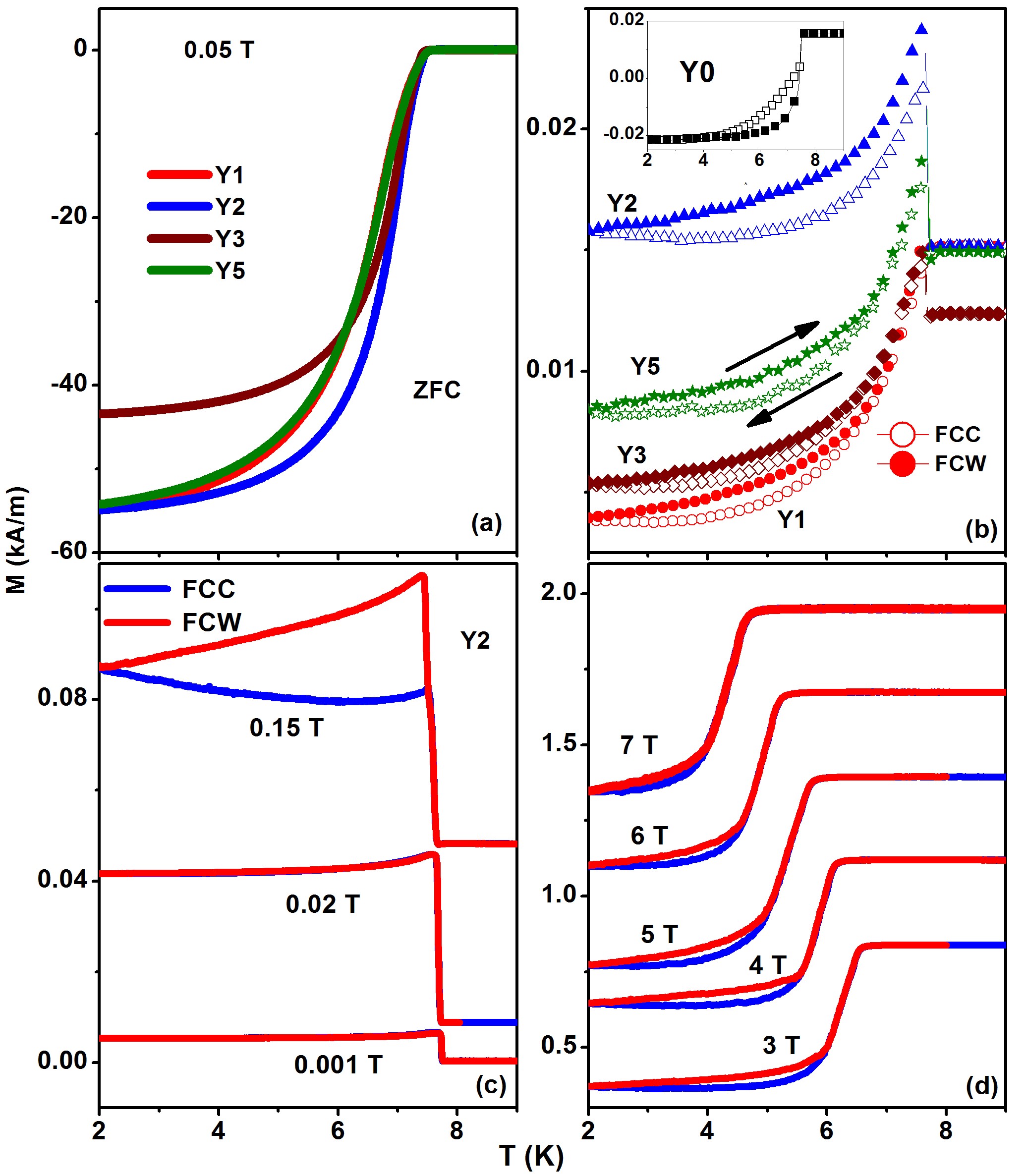}
  \caption{a) ZFC magnetisation curves for Y1-Y5 alloys in presence of 0.05 T. (b) Magnetisation as a function of temperature in FCC and FCW protocols below $T_{c}$ for  alloys Y1-Y5 in presence of 0.05 T. The inset to this figure shows the same for Y0 alloy. The open and solid symbols represents FCC and FCW data points, respectively. The FCW magnetisation values are higher than FCC in Y1-Y5, which is not the case in Y0. (c) and (d) shows the FC M(T) curves in the field range 0.001-7~T Y2 alloy. At higher fields ($>$0.02 T), the magnetisation values obtained in FCW protocol is more than FCC.}
  \label{4} 
\end{figure}

Fig.\ref{4}(a) shows the ZFC magnetisation of the Y1-Y5 alloys in presence of 0.05 T. All the alloys show a diamganetic behavior below the $T_{c}$, as is usually seen in any conventional type-II superconductor. Fig. \ref{4}(b) shows the temperature dependence of magnetisation ($M$) in the FCC {$M_{FCC}$}) and FCW($M_{FCW}$) protocols in the presence of 0.05 T field for Y1-Y5 alloys, and the same for Y0 is shown in the inset to fig.\ref{4}(b). In case of Y0, \textit{$M_{FCC}$} sharply reduces to a negative value when temperature is reduced below $T_{c}$, and attains a negative saturation value for $\sim$\textit{T}$<$$<$\textit{T}$_{c}$. On contrary, apart from the increasing magnetization for Y1-Y5 alloys just below T$_{c}$, we have observed a switching  over off the FCC and FCW curves. The higher positive values of M$_{FCW}$ than M$_{FCC}$ in Y1-Y5 alloys, is a manifestation of the HFPME. The FCC and FCW curve reversal in Y1-Y5 alloys is observed for fields above 0.02 T. Even though we observe clear paramagnetic response below 0.02 T, we do not find swapping of FCC and FCW curves. Hence, we associate the behavior at fields above 0.02~T with HFPME .

\par
\begin{figure}[t]
\centering
  \includegraphics[width=\columnwidth]{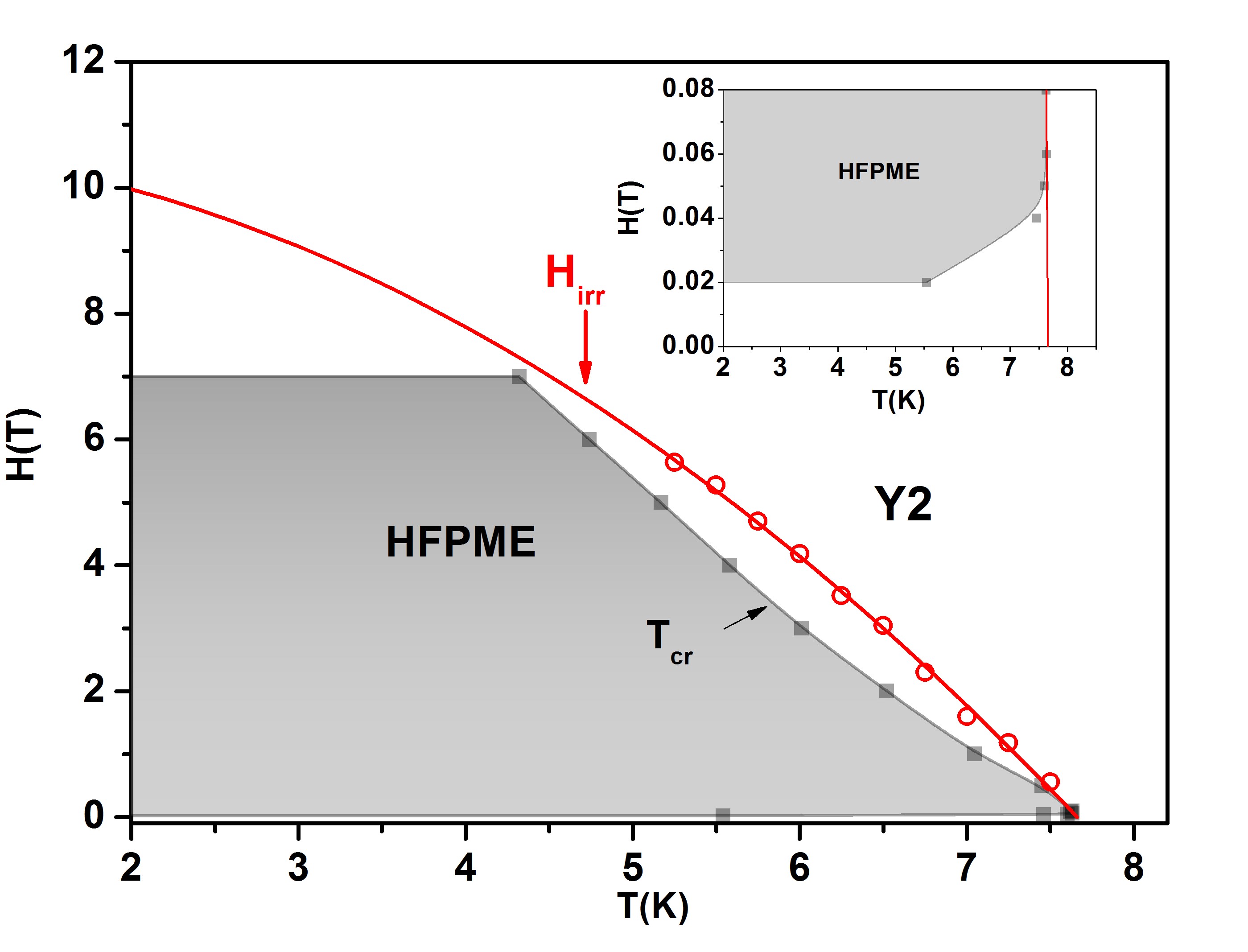}
  \caption{Phase diagram depicting the region of occurrence of HFPME in Y2 alloy. in mixed state. The solid line for H$_{irr}$ curve is a fit using the equation H$_{irr}$(T) = H$_{irr}$(0)[1-${T/T_{c}}^2$]. The inset to the figure indicates the absence of HFPME below 0.02 T.}
  \label{5}
\end{figure}

 In Y2 sample, the FC magnetisation curves obtained in 0.05 T field remain above the normal state magnetisation at all temperature below $T_{c}$ (fig.\ref{4}(b)). This indicates the HFPME is much prominent in Y2. Hence we measured the temperature dependence of magnetisation of Y2 sample in various applied magnetic fields. Fig.\ref{4}(c) and \ref{4}(d) illustrates the change in nature of FCC and FCW curves when field is increased from 1 mT to 7~T. In the presence of 1 mT field, the FCC and FCW curves overlap in the entire temperature range (see \ref{4}(c)). On increasing the field to 20 mT, the M$_{FCW}$(T) curves below 5K goes above the M$_{FCC}$(T). This marks the emergence of HFPME in the alloy. The temperature below which this crossover in M  is observed, defined as crossover temperature (T$_{cr}$). For fields equal to or greater than 0.02 T, the magnetization values during FCC is higher than that during FCW below T$_{cr}$. The $T_{cr}$ in the field range 0.03-0.1 T lies just below the $T_{c}$. However, for H~$>$ 0.1 T fields, significant difference between $T_{cr}$ and $T_{c}$ is observed. The $T_{cr}$ shifts towards lower temperature with increasing field. The hysteresis between FCC and FCW curves in the M axis, initially increases with increase in field upto 4 T, and then gradually decreases. Such irreversibility is observed upto the maximum field measured (7 T) for the Y2 alloy. However, it disappears in the Y1, Y3, Y5 alloy at 7 T. Fig. \ref{5} shows the crossover temperature obtained in various fields, indicating the region where HFPME prevails in Y2. The HFPME region  lies just below the $H_{irr}$ line, depicting the association of HFPME with flux pinning. The  characteristic fields $H_{irr}$ is acquired from the J$_{c}$(H) plots, as the field below which the J$_{c}$ drops sharply.

\par
\begin{figure}[t]
\centering
  \includegraphics[width=\columnwidth]{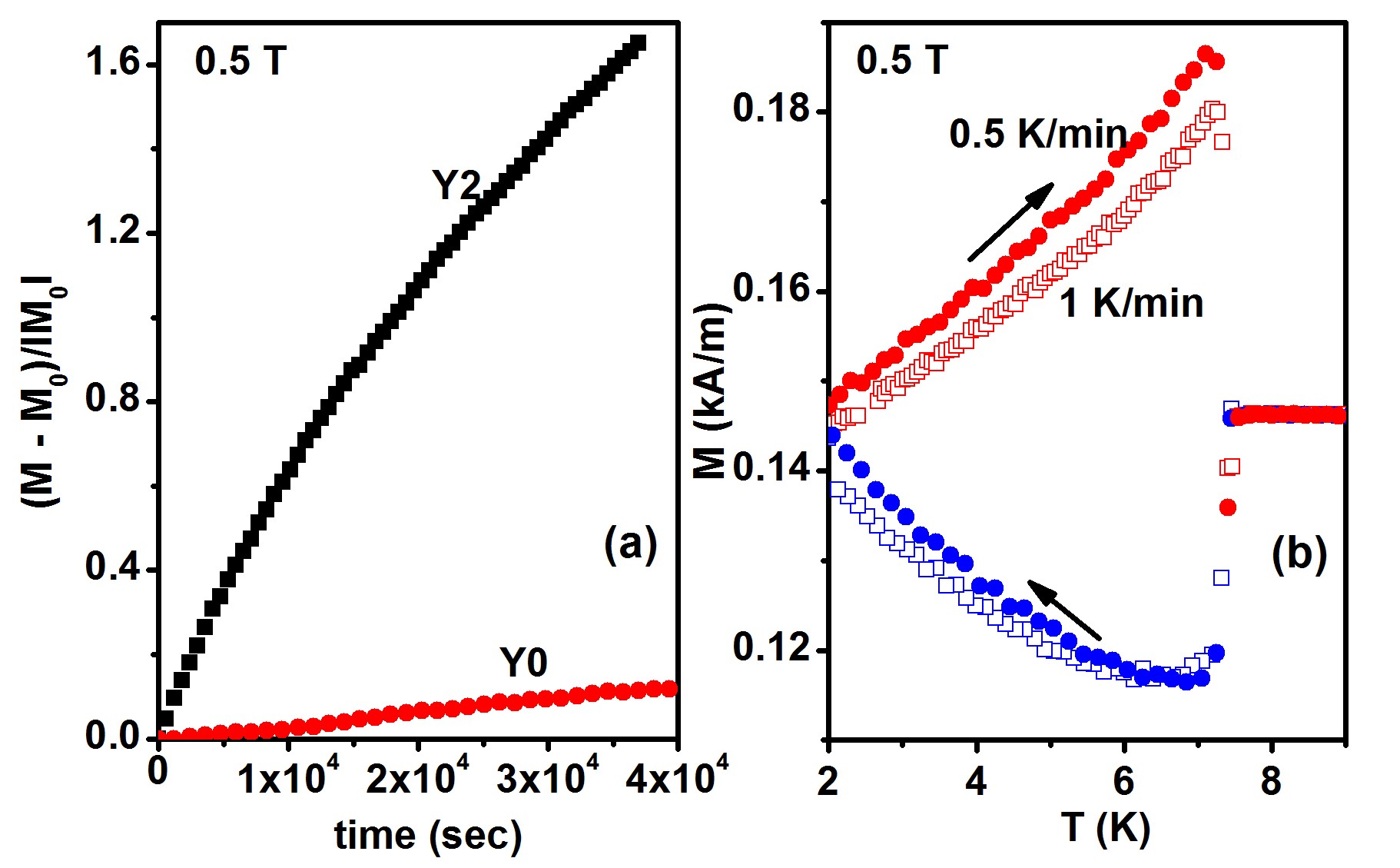}
  \caption{(a)Relaxation measurements depicting a strong time dependence of FCW magnetisation value measured at 0.5 T at 4 K. The magnetisation values become progressively more positive for Y2 as well as Y0. However, the relaxation in Y0 is almost negligible relative to Y2.(b) Temperature sweep rate dependence of FCC and FCW curves in presence of 0.5 T for Y2. For slower rate, the irreversibility is larger. This indicates that flux creep plays an active role in HFPME}
  \label{6}
\end{figure}

 Apart from higher magnetisation value in FCW than in FCC, the relaxation of $M$ values towards more positive value is a signature of HFPME. This paramagnetic relaxation has been observed in all samples exhibiting HFPME. Fig.\ref{6}(a) shows increase in $M_{FCW}$ with time in the presence of 0.5 T at 4 K, for Y0 and Y2 alloys.  M$_{0}$ represents the M value just before relaxation measurements were started. In a time period of 40,000 sec, the M increases by almost 160$\%$ and 10$\%$ in case of Y2 and Y0, respectively. This indicates that the relaxation in Y0 alloy is insignificant as compared to Y2 alloy. Similar effect has also been observed in other Y-containing alloys. This confirms the fact that relaxation is an characteristic feature of samples exhibiting HFPME.

 Fig. \ref{6}(b) shows the temperature dependent magnetisation curves performed at different sweep rates in the presence of 0.5 T field for the Y2 alloy. The FCC magnetisation value attains a more positive  for the slower sweep rate of 0.5 K/min, as compared to 1 K/min sweep rate. These observations also rule out the possibility of inhomogeneous cooling of sample as a reason for HFPME. Moreover, the amount of hysteresis, varies inversely to the sweep rates. 
%(The measurements performed at a slower rate allows more number of flux lines that are thermally de-pinned from weak pinning centers (grain boundaries here), to creep into secondary phase non-superconducting yttrium precipitates.)%

 We have shown that flux creep in Y2 alloy is much more significant than Y0, which could possibly a reason behind observation of HFPME. However, flux creep plays an associative role in bringing out the HFPME. The presence of an inhomogeneous flux-line pinning in a sample, is a necessary condition to observe this effect \cite{matin2013high,matin2015high}.  Such scenario becomes possible when different kinds of pinning centres are present, that can effectively pin the flux lines in various fields (that have different pinning efficiencies in various fields).

%\section{Discussion}

The HFPME has been reported in Ti rich V-Ti alloys, namely V$_{0.2}$Ti$_{0.8}$ and V$_{0.3}$Ti$_{0.7}$, in the field range 0.05 - 1.5 T\cite{matin2015high,matin2013high}. The $G_{i}$ number in these alloys is about $10^{-5}$\cite{matin2015high}, which is quite higher than any low T$_{c}$ conventional superconductor. This indicates that flux creep is significant in these alloys. The presence of $\alpha^{'}$ phase in V$_{0.2}$Ti$_{0.8}$ along the edge of sample and  presence of needle shaped $\alpha$ phase throughout the V$_{0.3}$Ti$_{0.7}$ sample, helped in creating a spatial inhomogeneity in flux pinning. It was also found that the presence of $\alpha $ phase over the V$_{0.6}$Ti$_{0.4}$ was not enough for observation of HFPME, due to negligible flux creep \cite{matin2015high}. The difference in the observation of HFPME in all the V-Ti-Y alloys as compared to the V-Ti alloys, is the range in which the HFPME is seen. This clearly indicates the source of inhomogeneous flux line pinning in V-Ti-Y alloys is different from V-Ti alloys.

\par
\begin{figure}
\centering
  
  \includegraphics[width=\columnwidth]{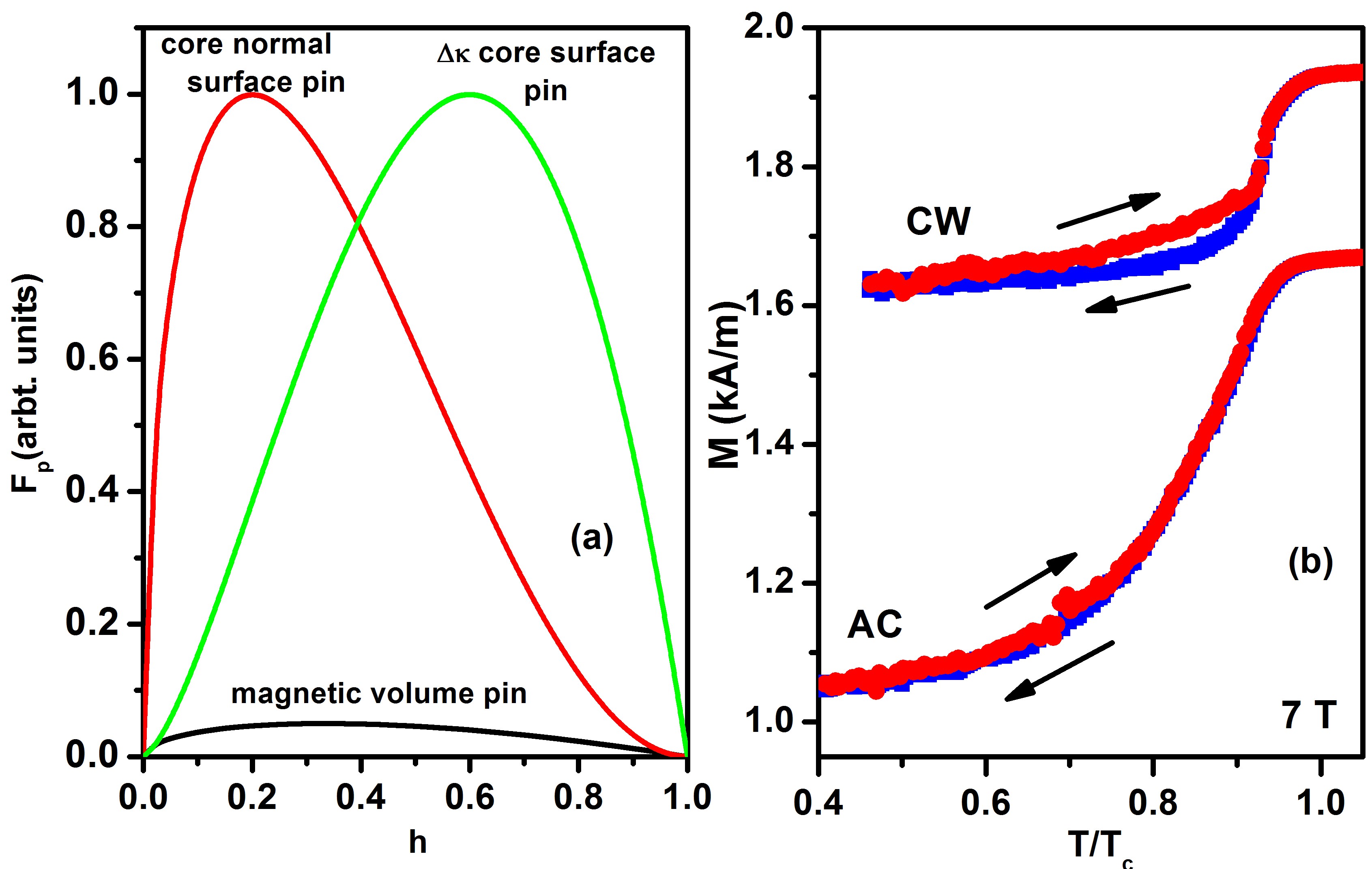}
\caption{a)The relative strengths of various types of pinning centres for $\kappa$ = 25. Clearly, it shows the weak pinning strength of magnetic volume pins as compared to core surface pins for a high $\kappa$ superconductor.(b) A comparison of M$_{FC}$(T) in presence of 7 T for as cast and cold-worked Y3 sample. The cold-worked samples shows the presence of HFPME, which is absent in its as cast counterpart. }
\label{7}
\end{figure}

 We have found that the addition of yttrium increases the  grain boundary density and dislocations in the alloys. These  grain boundaries and dislocations in the yttrium containing alloys pin the flux lines in the low and high magnetic fields respectively \cite{ramjan}. At low magnetic fields, the flux lines creep from dislocations to grain boundaries due to the fact that the pinning strength (F$_{p}$) of dislocations is significantly lower than that of grain boundaries at low H. On the other hand, F$_{p}$ of dislocations is higher for H approaching H$_{irr}$ as compared to that of grain boundaries. Hence, in such situations the flux lines creep from grain boundaries to dislocations. This creates higher flux density at the high strength pinning centres at a given H resulting in an inhomogeneous pinning of flux lines in these sample. The qualitative pinning force function of grain boundaries (core normal surface pins) and dislocations (core $\Delta\kappa$ surface pins) is shown in fig.\ref{7}(a) \cite{dewhughes, matin2015critical}. In addition, the flux lines can also be pinned at the precipitate interface. The size of precipitates in V$_{0.6}$Ti$_{0.4}$-Y alloys are larger as compared to the penetration depth ($\lambda$) \cite{ramjan}, and in such cases flux pinning is via magnetic interaction \cite{dewhughes,campbell1968pinning,coote1972flux}. The strength of flux pinning via magnetic interaction is 1/$\kappa$ times that of pinning by core interaction \cite{dewhughes, campbell1968pinning}.  V$_{0.6}$Ti$_{0.4}$-Y being a high $\kappa$ superconductor ($\kappa = 25$) \cite{ramjan}, the pinning via magnetic interaction is quite low compared to core pinning (see fig.\ref{7}(a). Clearly, the magnetic volume pins are the weakest pinning centres and hence do not contribute to enhancement of J$_{c}$ in V$_{0.6}$Ti$_{0.4}$-Y alloys. The very low pinning strength of Y-rich precipitates further encourages the inhomogeneous flux line pinning in these alloys. 

\begin{table}[htb]
 \caption{Pinning function of $F_{p}(h)$ associated with various types and geometry of pinning centres.}
 \centering % used for centering table
 \begin{tabular}{c c} % centered columns (4 columns)
 \hline\hline %inserts double horizontal lines
 Type & Functional form \\ % inserts table
 %of pins & type &  form &  of maximum \\ [0.8ex] 
 %heading
 \hline % inserts single horizontal line
 Magnetic volume & $ h^{0.5} (1-h)/\kappa^{3} $  \\ % inserting body of the table
 Core normal surface & $ h^{0.5} (1-h)^2/\kappa^{2}  $  \\
 Core $\Delta \kappa$& $ h^{1.5} (1-h)/\kappa^{2} $ \\ [1ex] % [1ex] adds vertical space
 \hline\hline %inserts single line
 \end{tabular}
 \label{II} % is used to refer this table in the text
 \end{table}

\par The presence of various pinning centres with different pinning efficiencies can possibly explain, the HFPME effect observed in our samples. As per Meissner effect while cooling down the sample below T$_{c}$, the flux lines are expelled away from sample in homogeneous superconductors. However, this flux lines can either be expelled out of the samples, or towards the non-superconducting regions in a non-homogeneous superconductors. The flux lines instead of being expelled out completely, can be pinned by the non-superconducting precipitates, grain boundaries or dislocations present in these alloys. The yttrium precipitates being very large, provides much interfacial area and hence can accommodate multiple flux lines. However, the pinning strength of the precipitates is very low. Due to thermal energy, the flux lines get de-pinned easily from the precipitates and may creep out of the pins. However, the precipitates are mainly present along the grain boundaries, which act as a strong pinning centres. Hence, the de-pinned flux lines can get pinned at the grain boundaries or at the dislocations present in the samples. This creates higher flux density at the grain boundaries and the dislocations, resulting in an inhomogeneous pinning of flux lines in these sample. Moreover, a few portions of the sample (like the region surrounding the precipitates) the flux density is much lower than the equilibrium distribution. The depletion of flux lines can facilitate entry of additional flux lines, leading to a paramagnetic response, and hence the HFPME is observed.

\par Increasing the number of dislocations can allow more number of flux lines, that get de-pinned from weak pinning centres (precipitates and grain boundaries in this field regime) to get pinned. This might increase the range in which the HFPME effect is observed. In order to verify this fact, we measured the magnetisation of a cold-worked (CW) sample in 7 T. It is well-known that, cold working increases the number of dislocations present in a sample. Fig.\ref{7}(b) shows a comparison of the FC magnetisation curves in 7 T for the as-cast (AC) and CW samples of Y3. It can be clearly seen that HFPME is absent in the Y3 AC sample, since FCC and FCW curves almost overlap. However, on cold-working the HFPME appears. This shows that HFPME can be introduced by simply cold-working in these samples. 
%   (\textbf{Note:: The explanation of HFPME considering the difference in pinning strength of grain boundaries and dislocation also seems plausible. However, pinning strength of dislocations at low field is quite high as compared to precipitates. The pinning strength of grain boundaries and dislocation varies by less than an order of magnitude only. Moreover, if this was true, we must observe HFPME in every system with large number of dislocations and higher grain boundary density}) 

\par One of the most distinct outcome of this work, is the observation of HFPME upto a magnitude of 7~T. Only YBa$_{2}$Cu$_{3}$O$_{7-\delta}$ high $T_{c}$ superconductor shows HFPME in fields above 7 T. Recent studies suggests that odd-frequency pairing can be present at the superconductor-normal(S/N) interface, even in a s-wave superconductor\cite{tanaka2008odd}.The presence of odd-frequency pairing can even give rise to paramagnetic response\cite{di2015intrinsic}. Since, large number of S/N interfaces is present in V-Ti-Re alloys, it would be interesting to check whether the paramagnetic response is due to the proximity induced odd-frequency pairing at S/N interface.

\section{Conclusion}
To conclude, our temperature and time dependent magnetisation measurements indicate the existence of HFPME in all the yttrium substituted samples in the range of 0.04-7 T, which is in contrast to the parent Y0 alloy where HFPME is absent. The grain boundaries, dislocations and yttrium precipitates of size larger than penetration depth are the important microscopic features found in these alloys. The precipitates are the weakest pinning centres in the entire field regime, in these alloys. Relatively, the pinning strength of grain boundaries and dislocations is higher. The huge difference in pinning strength of various pinning centres in a particular field, is the origin of HFPME in such alloys. Our results also indicate that, HFPME is most distinguished in sample with the most number of weak as well as strong pinning centres. The range in which HFPME is observed can be extended by simply creating high field strong pinning centres.


\section*{REFERENCES}
\begin{thebibliography}{1}
%\biboptions{square}

\bibitem{prekul1975effect}
Prekul AF, Rassokhin VA and Volkenshteln NV 1974 \textit{Sovt. J.  Exp. and Theo. Phys.} \textbf{67} 1134

%2
\bibitem{matin2014influence}
Matin Md, Chandra LS Sharath, Pandey S K, Chattopadhyay MK, Roy SB 2014 \textit{The Eur. Phys. J. B} \textbf{87} 6

%3
\bibitem{matin2014spin}
Matin Md, Chandra LS Sharath, Meena Radhakishan, Chattopadhyay MK, Sinha AK, Singh MN and Roy SB 2014 \textit{Phys. B.  Cond. Matt.} 20-25 \textbf{436}

%4
\bibitem{paul2019renormalization}
Paul Sabyasachi, Chandra, LS Sharath, Chattopadhyay MK 2019 \textit{J. Phys. Cond. Matt.} \textbf{31} 475801
%5
\bibitem{tai2007superconducting}
Tai M, Inoue K, Kikuchi A, Takeuchi T, Kiyoshi T, Hishinuma Y 2007 \textit{IEEE. Tran. App. Super.} \textbf{17} 2542-2545

\bibitem{bellin1970critical}
Bellin PH, Gatos HC, Sadagopan V 1970 \textit{J. Appl. Phys.} \textbf{41} 2057-2059

\bibitem{matin2015high}
Matin Md, Chattopadhyay MK, Chandra LS Sharath and Roy SB 2015 \textit{Super. Sci.Tech} \textbf{29} 025003

\bibitem{matin2013high}
Matin Md, Chandra LS Sharath, Chattopadhyay MK, Singh MN, Sinha AK and Roy SB 2013 \textit{Super. Sci.Tech} \textbf{26} 115005

\bibitem{sundar2015high}
Sundar Shyam, Chattopadhyay MK, Chandra LS Sharath and Roy SB 2015 \textit{Super. Sci.Tech} \textbf{28} 075011

\bibitem{dias2016high}
Dias FT, Vieira Valdemar das Neves, Wolff-Fabris Frederik, Kampert Erik Gouv{\^e}a CP, Campos APC, Archanjo BS, Schaf Jacob, Obradors X, Puig T and others  2016 \textit{Physica C} \textbf{525} 105



\bibitem{dias2001paramagnetic}
Dias FT, Pureur P, Rodrigues Jr P and Obradors X 2001 \textit{Physica C} \textbf{354} 219

\bibitem{dias2004paramagnetic}
Dias FT, Pureur Paulo, Rodrigues Jr Pedro and Obradors Xavier 2004 \textit{Phys. Rev. B} \textbf{70} 224519

\bibitem{terentiev1999paramagnetic}
Terentiev A, Watkins DB, De Long LE, Morgan DJ and Ketterson JB 1999 \textit{Phys. Rev. B} \textbf{60} R761

\bibitem{prokhorov2009flux}
Prokhorov VG,Svetchnikov VL, Park JS, Kim GH, Lee YP, Kang JH, Khokhlov VA and Mikheenko P 2009 \textit{Super. Sci.Tech} \textbf{22} 045027

\bibitem{kostic1996paramagnetic}
Kosti{\'c} P, Veal B, Paulikas AP, Welp U, Todt VR, Gu, Geiser U, William JM, Carlson  KD, Klemm RA 1996 \textit{Phys. Rev. B} \textbf{53} 791

\bibitem{matin2015critical}
Matin Md and Chandra LS Sharath and Chattopadhyay MK and Meena RK and Kaul Rakesh and Singh MN and Sinha AK and Roy SB 2015 \textit{Physica C} \textbf{512} 32

\bibitem{chu2006extrinsic}
Chu Shaoyan, Schwartz Adam J, Massalski Thaddeus B and Laughlin David E \textit{Appl. Phys. Lett} \textbf{89} 111903

\bibitem{ge2017paramagnetic}
  Ge Jun-Yi, Gladilin Vladimir N, Sluchanko Nikolay, Lyashenko A, Filipov Volodimir B, Indekeu Joseph O and Moshchalkov Victor V 2017 \textit{New. Jour. Phys.}\textbf{19} 093020
\bibitem{svedlindh1989anti}
 Svedlindh P, Niskanen K,Norling P,Nordblad P, Lundgren L, L{\"o}nnberg B and Lundstr{\"o}m T 1989 \textit{Physica C} \textbf{162} 1365
 

\bibitem{dewhughes}
Dew-Hughes D 1974 \textit{The. Phil. Mag.} \textbf{30} 293


\bibitem{tanaka2008odd}
Tanaka Yukio and Golubov Alexander A 2008 \textit{Physica C} \textbf{468} 589

\bibitem{di2015intrinsic}
Di Bernardo A, Salman Z, Wang XL, Amado M, Egilmez M, Flokstra M G, Suter A,Lee Steve L, Zhao JH, Prokscha T and others

\bibitem{bose2008electron}
Bose S K 2009 \textit{J. Phys. Cond. Matt.} \textbf{21} 025602

\bibitem{chu2006extrinsic}
Chu Shaoyan, Schwartz Adam J, Massalski Thaddeus B and Laughlin David E \textit{Appl. Phys. Lett} \textbf{89} 111903

\bibitem{matin2013magnetic}
Matin Md, Sharath Chandra LS, Chattopadhyay MK, Meena RK, Kaul Rakesh, Singh MN, Sinha AK and Roy SB 2013 \textit{J. Appl. Phys.} \textbf{113} 163903

\bibitem{ge2017paramagnetic}
  Ge Jun-Yi, Gladilin Vladimir N, Sluchanko Nikolay, Lyashenko A, Filipov Volodimir B, Indekeu Joseph O and Moshchalkov Victor V 2017 \textit{New. Jour. Phys.}\textbf{19} 093020

\bibitem{svedlindh1989anti}
 Svedlindh P, Niskanen K,Norling P,Nordblad P, Lundgren L, L{\"o}nnberg B and Lundstr{\"o}m T 1989 \textit{Physica C} \textbf{162} 1365
 
\bibitem{paul2021grain}
Paul S, Ramjan SK, Venkatesh R, Chandra LS Sharath and Chattopadhyay MK

\bibitem{ramjan}
Ramjan SK, Chandra LS Sharath, Singh Rashmi and Chattopadhyay MK (unpublished) and reference therein

\bibitem{hamlin2006superconductivity}
Hamlin JJ, Tissen VG and Schilling JS 2006 \textit{Phys. Rev. B} \textbf{73} 

\bibitem{chan2010thermodynamic}
Chan Wren, Gao Michael C and Do{\u{g}}an {\"O}mer N and King Paul 2010 \textit{J. Phase. Equilib. Diff. } \textbf{31} 425-432 

\bibitem{campbell1968pinning}
Campbell AM, Evetts JE and Dew-Hughes D 1968 \textit{Phil. Magaz.} \textbf{18} 152 
  
\bibitem{coote1972flux}
Coote RI, Evetts JE and Campbell AM 1972 \textit{Can. J. Phys.} \textbf{50} 5
  




 



 
\end{thebibliography}
\end{document}